\documentclass[sigconf, screen]{acmart}

\AtBeginDocument{%
  }

\copyrightyear{2025}
\acmYear{2025}
\setcopyright{cc}
\setcctype{by-nc-sa}
\acmConference{Access InContext Workshop @ CHI’25}{April 26}{Yokohama, Japan}
\acmBooktitle{Access InContext Workshop @ CHI’25, April 26, 2025, Yokohama, Japan}

\acmDOI{}                          

\acmISBN{}                         
\acmPrice{}                        

\begin{document}

\title{Human-in-the-Loop Optimization for Inclusive Design: \\Balancing Automation and Designer Expertise}

\author{Pascal Jansen}
\email{pascal.jansen@uni-ulm.de}
\orcid{0000-0002-9335-5462}
\affiliation{%
  \institution{Institute of Media Informatics, Ulm University}
  \city{Ulm}
  \country{Germany}
}

\renewcommand{\shortauthors}{Pascal Jansen}

\begin{abstract}
Accessible and inclusive design has gained increased attention in HCI, yet practical implementation remains challenging due to resource-intensive prototyping methods. Traditional approaches such as workshops, A/B tests, and co-design sessions struggle to capture the diverse and complex needs of users with disabilities at scale. This position paper argues for an automated, accessible Human-in-the-Loop (HITL) design optimization process that shifts the designer’s role from directly crafting prototypes to curating constraints for algorithmic exploration. By pre-constraining the design space based on specific user interaction needs, integrating adaptive multi-modal feedback channels, and personalizing feedback prompts, the HITL approach could efficiently refine design parameters, such as text size, color contrast, layout, and interaction modalities, to achieve \textit{optimal} accessibility. This approach promises scalable, individualized design solutions while raising critical questions about constraint curation, transparency, user agency, and ethical considerations, making it essential to discuss and refine these ideas collaboratively at the workshop.
\end{abstract}




\maketitle

\section{Introduction}
\label{sec:introduction}
Accessible and inclusive design has gained significant attention in HCI, evidenced by a steady increase in accessibility-related publications at venues like CHI~\cite{mack2021we}. Disability rights advocates consistently call for the direct involvement of people with disabilities throughout the design process~\cite{hofmann2020living,mankoff2010disability,spiel2020nothing}. Yet, achieving practical inclusivity remains a challenge.

Current design and prototyping of user interfaces primarily rely on resource-intensive methods such as workshops, A/B tests, and co-design sessions. While these methods can yield valuable insights, they are time-consuming and challenging to scale to individuals with diverse needs. Even well-intentioned design teams struggle to keep pace with rapidly evolving accessibility guidelines, device capabilities, and user preferences, making it challenging to capture the complexity of diverse disabilities~\cite{mack2022anticipate,spiel2020nothing}. 

To overcome these limitations, adaptive techniques known as Human-in-the-Loop (HITL) optimization have emerged within HCI. These approaches leverage algorithms, such as Bayesian Optimization (BO), to iteratively refine designs based on user feedback~\cite{chiu2020human,koyama2020sequential,zhong2021spacewalker}. In a HITL process, BO automatically explores possible designs by adjusting design parameter values such as text size, color contrast, layout arrangements, or interaction modalities. Each design is evaluated by a user who provides feedback as subjective assessments of user experience~\cite{jansen2025opticarvis,colley2025improving} or performance metrics such as task completion time or error rate~\cite{chan2022bo}. The optimizer treats this feedback as data points that inform a statistical model. Notably, BO is efficient even without prior data, enabling it to start optimizing immediately \cite{brochu2010tutorial}. BO specifically balances exploring new design variations (\textit{exploration}) with refining promising solutions based on prior user feedback (\textit{exploitation})~\cite{brochu2010tutorial,borji2013bayesian,chong2021interactive}. Unlike traditional methods like A/B tests, which evaluate a limited number of design variants, BO efficiently searches for an \emph{optimal} design, that is, a design parameter combination that best aligns with multiple design objectives~\cite{marler_survey_2004}, such as usability or accessibility. Thus, BO can identify effective designs in a few iterations~\cite{brochu2010tutorial}, making it well-suited for adaptive prototyping.

However, current HITL optimization approaches often assume uniform user feedback channels, such as text-based surveys or numerical rating scales, excluding participants who face barriers using these modalities~\cite{bennett2016using,brewer2018facilitating}. For example, many digital prototyping tools still lack compatibility with screen readers or fail to support multi-modal interactions (e.g., combining visual, auditory, and tactile inputs), resulting in significant accessibility gaps~\cite{li2021accessibility}. Consequently, gathering meaningful feedback from users with disabilities becomes challenging, typically requiring additional manual processes that are labor-intensive and may inadvertently overlook crucial user insights into designs.

This position paper advocates replacing traditional, manually driven prototyping with an automated, accessible HITL optimization process. Here, the designers shift from directly creating prototypes to curating design constraints that guide the optimizer. This approach promises a scalable, adaptive process that moves beyond “one-size-fits-most” solutions toward designs that respond dynamically to users' diverse preferences and needs.

\section{Future Perspectives: Automating Inclusive Design Through Accessible Human-in-the-Loop Optimization}
\label{sec:future_perspectives}
Automating inclusive design through HITL optimization requires carefully balancing automation, designer control, and individual user requirements. The following concepts outline this approach:

\paragraph{\textbf{Pre-Constraining the Design Space by User Needs}}
Traditionally, designers define a static set of design parameters, such as text size or color schemes, used for all users. However, simultaneously including all possible parameters would significantly increase complexity and require extensive iterations in a HITL optimization. An alternative is to use an \textit{accessible HITL optimization} approach, where designers could pre-define constraints tailored to specific interaction needs relevant to particular user groups. For instance, a constraint may specify that visually impaired users benefit from enlarged text sizes and high-contrast color schemes \cite{zhao2019seeingvr}. In contrast, such a constraint could limit or exclude high-brightness parameters for users sensitive to bright lighting in virtual environments \cite{viirre2002direct}. By defining constraints tailored to the needs of specific user groups, the optimizer could more efficiently explore the reduced design space, rapidly converging on an \textit{optimal} design \textbf{per} user.

\paragraph{\textbf{Accessible Feedback Channels}}
Conventional HITL approaches typically rely on standardized text-based surveys or rating scales, inherently excluding users with sensory, cognitive, or motor impairments~\cite{bennett2016using,brewer2018facilitating}. Accessible HITL optimization could address this limitation by enabling multiple accessible feedback methods (e.g., voice input, tactile responses, or automatically detected behaviors). For example, a visually impaired user may provide spoken responses rather than rating interfaces visually, while a user with motor impairments might give tactile feedback via simplified haptic devices. Additionally, if the system detects repeated navigation errors (such as cursor misplacement), it can dynamically switch from visual prompts to voice or haptic interactions. This flexibility would ensure all users can provide meaningful feedback, improving the accuracy and inclusivity of the optimization process.

\paragraph{\textbf{Personalized Feedback Prompting}}
HITL feedback prompts' timing, frequency, and complexity should be tailored individually. Typical HITL systems often assume all users can equally engage with uniform feedback prompts, neglecting user communication styles and cognitive capacity differences \cite{knauper1997question,schneider2024can}. Accessible HITL optimization could adapt feedback prompts to individual preferences and abilities. For example, the optimizer may issue shorter and more frequent feedback prompts to users with cognitive processing differences while providing fewer but longer feedback forms for users comfortable engaging in more detailed reflection. By personalizing feedback prompting, the system would retrieve higher-quality data (e.g., being more representative \cite{knauper1997question}), enhancing the quality of subsequent design adaptations.

\section{Opportunities and Challenges}
\label{sec:opportunities_challenges}
Accessible HITL optimization through (1) pre-constrained design spaces, (2) multi-modal feedback channels, and (3) personalized feedback prompting could enable scalable and individualized inclusive design. Pre-defining constraints (e.g., enforcing high-contrast modes for visually impaired individuals) could enable the optimizer to explore a focused design space efficiently, resulting in designs that better meet specific user needs.

However, this approach also faces challenges. Defining appropriate constraints that capture the diverse needs of all users is complex; overly narrow constraints might exclude some users, while overly broad ones may slow optimization. Additionally, automated systems risk misinterpreting subtle user behaviors. For example, repeated webpage zooming may indicate a usability issue or be a personal browsing habit. To mitigate such ambiguities, it is essential to implement mechanisms that allow users to override or refine automated decisions. User trust is another critical concern; transparent, accessible explanations of automated decisions (e.g., via simplified visual or auditory narratives) are necessary. Finally, resolving conflicting accessibility preferences across large user populations remains an ongoing challenge that may require hybrid approaches integrating automated optimization with human oversight.

\section{Conclusion}
\label{sec:conclusion}
This position paper argues for shifting the designer's role from directly creating individual prototypes toward strategically defining constraints for automated and accessible HITL optimization. By pre-constraining the design space according to specific user interaction needs, incorporating dynamically adaptive multi-modal feedback channels, and personalizing feedback collection, this approach promises scalable and individualized accessibility in prototyping.

Key discussion topics for the workshop include the feasibility of curating effective constraints, the design and ethics of adaptive feedback modalities, methods to preserve user agency, and strategies for maintaining transparency and resolving conflicting user needs. Interdisciplinary collaboration is essential to address these challenges and advance inclusive, automated prototyping methodologies. The workshop offers an ideal venue for exploring these issues and collectively shaping the future of accessible design.



\bibliographystyle{ACM-Reference-Format}
\bibliography{chi25-workshop-submission}


\end{document}